\documentclass[12pt,preprint]{aastex}

\newdimen\minuswidth    %define @ width of minus sign for tables
\setbox0=\hbox{$-$}
\minuswidth=\wd0
\catcode`@=\active
\def@{\kern\minuswidth}
\newdimen\digitwidth    %define ! a one digit width for tables
\setbox0=\hbox{\rm0}
\digitwidth=\wd0
\catcode`!=\active
\def!{\kern\digitwidth}

\begin{document}
\shorttitle{High resolution infrared spectroscopy of NGC~6791}
\shortauthors{Origlia et al.}

\title{High resolution infrared spectroscopy of the old open cluster NGC~6791}
%\altaffillmark{1}
\altaffiltext{1}{
Data presented herein were obtained
at the W.M.Keck Observatory, which is operated as a scientific partnership
among the California Institute of Technology, the University of California,
and the National Aeronautics and Space Administration.
The Observatory was made possible by the generous financial support of the
W.M. Keck Foundation.}

\author{Livia Origlia}
\affil{INAF -- Osservatorio Astronomico di Bologna,
Via Ranzani 1, I--40127 Bologna, Italy,\\
livia.origlia@bo.astro.it}
\author{Elena Valenti}
\affil{Dip. di Astronomia, Universit\`a degli Studi di Bologna, 
Via Ranzani 1, I--40127 Bologna, Italy,\\
INAF -- Osservatorio Astronomico di Bologna,\\
elena.valenti3@unibo.it}
\author{R. Michael Rich}
\affil{Physics and Astronomy Bldg,
430 Portola Plaza  Box 951547
Department of Physics and Astronomy, University of California
at Los Angeles, Los Angeles, CA 90095-1547\\
rmr@astro.ucla.edu}
\author{Francesco R. Ferraro}
\affil{Dip. di Astronomia, Universit\`a degli Studi di Bologna, 
Via Ranzani 1, I--40127 Bologna, Italy,\\
francesco.ferraro3@unibo.it}

%\medskip

\begin{abstract}

We report abundance analysis for 6 M giant members of the old open cluster
NGC~6791, based on infrared spectroscopy  $(1.5-1.8 ~\mu \rm m)$ at R=25,000,
using the NIRSPEC spectrograph at the Keck II telescope.
We find the iron abundance 
$\rm \langle [Fe/H] \rangle = +0.35 \pm 0.02$, confirming the super solar metallicity 
of this cluster derived from optical medium-high resolution spectroscopy.  
We also measure C, O and other alpha element abundances, 
finding roughly solar [$\rm alpha$/Fe] and 
$\rm \langle [C/Fe] \rangle = -0.35$.  Our approach constrains [O/Fe] especially
well, based on the measurement of a number of OH lines near 1.6~$\rm \mu m$; we find
[O/Fe]=$-0.07 \pm0.03$. 
The Solar alpha enhancement
is in contrast to the composition of similar stars in the Galactic bulge.
We also find
low $\rm ^{12}C/^{13}C\approx10$, confirming the
presence of extra-mixing processes during the red giant phase of evolution, 
up to super solar metallicities.

\end{abstract}


\section{Introduction}
\label{intro}
The open cluster NGC~6791 is currently believed to be one of the most massive,
metal\--rich and oldest stellar system. For this reason it has been the subject of 
many
photometric \citep{kin65,HC81,DGG92,ATT85,kal90,kalud92,gar94,mey93,kalruc93,
tripicco95,cha99,stet03,carney05} and spectroscopic 
\citep{fj93,pet98,friel02,wj03} 
investigations. Its populous color\--magnitude diagram (CMD) suggests a mass
${\geq}4000M_{\odot}$ \citep{kalud92}, and an age 
in the 6-12~Gyr range, as
inferred from both optical 
and IR photometry 
\citep[see e.g][]{kalud92,tripicco95,cha99,stet03,carney05},
dependent on the adopted reddening and metallicity. 
Estimates of the cluster reddening also cover
some range, from E(B\--V)=0.10 \citep{j84} to E(B\--V)=0.22 \citep{kin65}, with
a mean value of E(B\--V)=0.16 which is in excellent agreement with \citet{sch98}
extinction maps, which gives E(B\--V)=0.15 (see \S~2).
NGC~6791 is a relatively distant cluster, with a suggested distance modulus
(m-M)$_0$ ranging from 12.60 \citep{ATT85} to 13.6 \citep{HC81}.
NGC~6791 has also a peculiar white dwarf luminosity function, and the
metallicity of the cluster has some bearing on the explanation of the 
WD properties \citep{bed05, han05}.

\begin{deluxetable}{llll}
\tabletypesize{\scriptsize}
\tablecaption{Our sample of observed giant stars in NGC~6791. \label{tab1}}
\tablewidth{0pt}
\tablehead{
\colhead{Star} & 
\colhead{2MASS}&         
\colhead{RA (2000)}&
\colhead{DEC (2000)}
}
\startdata
%\multicolumn{4}{l}{}\\
\#1 & 19211606+3746462 & 19h~21m~16s &+37d~46'~27'' \\
\#2 & 19204971+3743426 & 19h~20m~50s &+37d~43'~43'' \\
\#3 & 19213390+3750202 & 19h~21m~34s &+37d~50'~20'' \\
\#4 & 19204635+3750228 & 19h~20m~46s &+37d~50'~23'' \\
\#5 & 19205510+3747162 & 19h~20m~55s &+37d~47'~16'' \\
\#6 & 19205338+3748282 & 19h~20m~53s &+37d~48'~28'' \\
\enddata                       
\end{deluxetable}

However, as reviewed by \citep{stet03,carney05}, chemical abundances are difficult
to measure in this moderately distant and reddened cluster. 
Its most luminous stars are relatively faint
and the combination of low effective temperature and high metallicity 
make high resolution optical spectra difficult to analyze. 
Thus, metallicity
estimates have been derived mainly using three different 
technique such as {\it 1)} photometric metallicity indicators 
\citep{j84,cat86}; {\it 2)} low\-- and moderate\--resolution spectroscopy
\citep{fj93,pet98,friel02,wj03}; and {\it 3)} model isochrones
\citep[][and reference therein]{stet03,carney05}. To summarize, the
metallicity proposed for this cluster is in the range 
+0.11--+0.44 dex. 
The first work at medium-high resolution is the one by
\citet{pet98}, who measured a sample of warm HB stars at R=20,000, finding 
an iron abundance [Fe/H]=+0.4$\pm$0.1, a modest 
(if any) $\alpha$-enhancement (within a factor of 2), and 
about solar [C/Fe]. 
Very recently, two other spectroscopic studies on clump and Red Giant Branch (RGB) stars, 
have been performed 
by \citet{car06} and \citet{gra06} finding  [Fe/H]=+0.39$\pm$0.01 and [Fe/H]=+0.47$\pm$0.04,
respectively. 
\citet{car06} also find about solar [$\alpha$/Fe], while \citet{gra06} find [O/Fe] depleted 
by a factor of 2 with respect to the solar value.

The use of IR
spectroscopy offers an interesting alternative to optical spectroscopy, as it is less sensitive to the 
blanketing effects and more suitable to study cool and metal rich stars than the optical
spectral range.   Our group has been undertaking a program using the NIRSPEC spectrograph \citep{ml98}
at Keck to obtain spectra of old metal rich stars in the bulge field \citep{ro05} and globular clusters 
\citep{ori03,ori04,or04,ori05} with the aim of studying 
the composition and chemical evolution of the bulge and globular
clusters.
Precise chemical abundances of NGC~6791 are 
crucial to constrain better the age of this cluster, which
deserves detailed investigations being one of the few examples in which we can 
study stars that formed very early in the evolution of the Galactic disk.
As underlined by \citet{carney05},
because NGC~6791 is both old and metal\--rich, it also plays a fundamental role 
in
calibrating several {\it "secondary"} metallicity indicators such as the low\--
to moderate\--resolution spectroscopy or photometry 
\citep[see e.g.][]{VFOa,VFOb}.  
In this context, we present high\--resolution IR spectra and the
abundance analysis of six bright giants in the open cluster NGC~6791. Our
observations, data reduction and abundance analysis follow in \S~2, 
while \S~3 discusses our results.
\begin{deluxetable}{lllllllllllll}
\tabletypesize{\scriptsize}
\tablecaption{Stellar parameters and chemical abundances for our sample of 
stars in NGC~6791. \label{tab2}}
\tablewidth{0pt}
\tablehead{
\colhead{Star} & 
\colhead{(J-K)$_0^a$}&         
\colhead{$\rm T_{eff}$}&
\colhead{log~g}&
\colhead{$\rm v_r^b$}&
\colhead{$\rm [Fe/H]$}& 
\colhead{$\rm [O/Fe]$}& 
\colhead{$\rm [Si/Fe]$}& 
\colhead{$\rm [Mg/Fe]$}& 
\colhead{$\rm [Ca/Fe]$}& 
\colhead{$\rm [Ti/Fe]$}& 
\colhead{$\rm [Al/Fe]$}& 
\colhead{$\rm [C/Fe]$} 
}
\startdata
%\multicolumn{13}{l}{}\\
\#1 & 1.13 &3600& 1.0 &-49  &+0.36  & -0.04  & -0.06  & -0.03  & +0.04  & +0.06  & +0.04  & -0.36 \\
   &&&&&$\pm$0.09  &$\pm$0.13  &$\pm$0.18  &$\pm$0.10  &$\pm$0.14  &$\pm$0.16  &$\pm$0.16  &$\pm$0.12\\
\#2 & 1.15 &3600& 1.0 &-52  &+0.33  & -0.05  & -0.03  & -0.03  & +0.07  & -0.03  & +0.07  & -0.33 \\
   &&&&&$\pm$0.09  &$\pm$0.13  &$\pm$0.18  &$\pm$0.10  &$\pm$0.14  &$\pm$0.16  &$\pm$0.16  &$\pm$0.12\\
\#3 & 1.02 &3800& 1.5 &-50  &+0.32  & -0.06  & +0.08  & -0.02  & +0.08  & +0.08  & +0.08  & -0.32 \\ 
   &&&&&$\pm$0.08  &$\pm$0.11  &$\pm$0.17  &$\pm$0.09  &$\pm$0.13  &$\pm$0.16  &$\pm$0.14  &$\pm$0.11\\
\#4 & 0.93 &4000& 1.5 &-50  &+0.38  & -0.08  & +0.08  & -0.05  & +0.02  & +0.02  & +0.02  & -0.38 \\
   &&&&&$\pm$0.08  &$\pm$0.09  &$\pm$0.21  &$\pm$0.08  &$\pm$0.13  &$\pm$0.14  &$\pm$0.14  &$\pm$0.11\\
\#5 & 0.92 &4000& 1.5 &-51  &+0.37  & -0.09  & +0.03  & -0.02  & +0.03  & +0.03  & +0.03  & -0.37 \\
   &&&&&$\pm$0.08  &$\pm$0.09  &$\pm$0.21  &$\pm$0.08  &$\pm$0.13  &$\pm$0.14  &$\pm$0.14  &$\pm$0.11\\
\#6 & 0.89 &4000& 1.5 &-52  &+0.36  & -0.11  & -0.01  & +0.00  & +0.04  & +0.04  & +0.04  & -0.36 \\
   &&&&&$\pm$0.07  &$\pm$0.09  &$\pm$0.21  &$\pm$0.08  &$\pm$0.12  &$\pm$0.14  &$\pm$0.13  &$\pm$0.10\\
\enddata                       
\tablenotetext{a}{
(J--K) colors are from 2MASS and have been corrected 
for reddening using \citet{sch98} extinction maps.} 
\tablenotetext{b}{Heliocentric radial velocity in $\rm km~s^{-1}$.}
\end{deluxetable}

\section{Observations and abundance analysis}

\begin{figure}
\epsscale{1}
\plotone{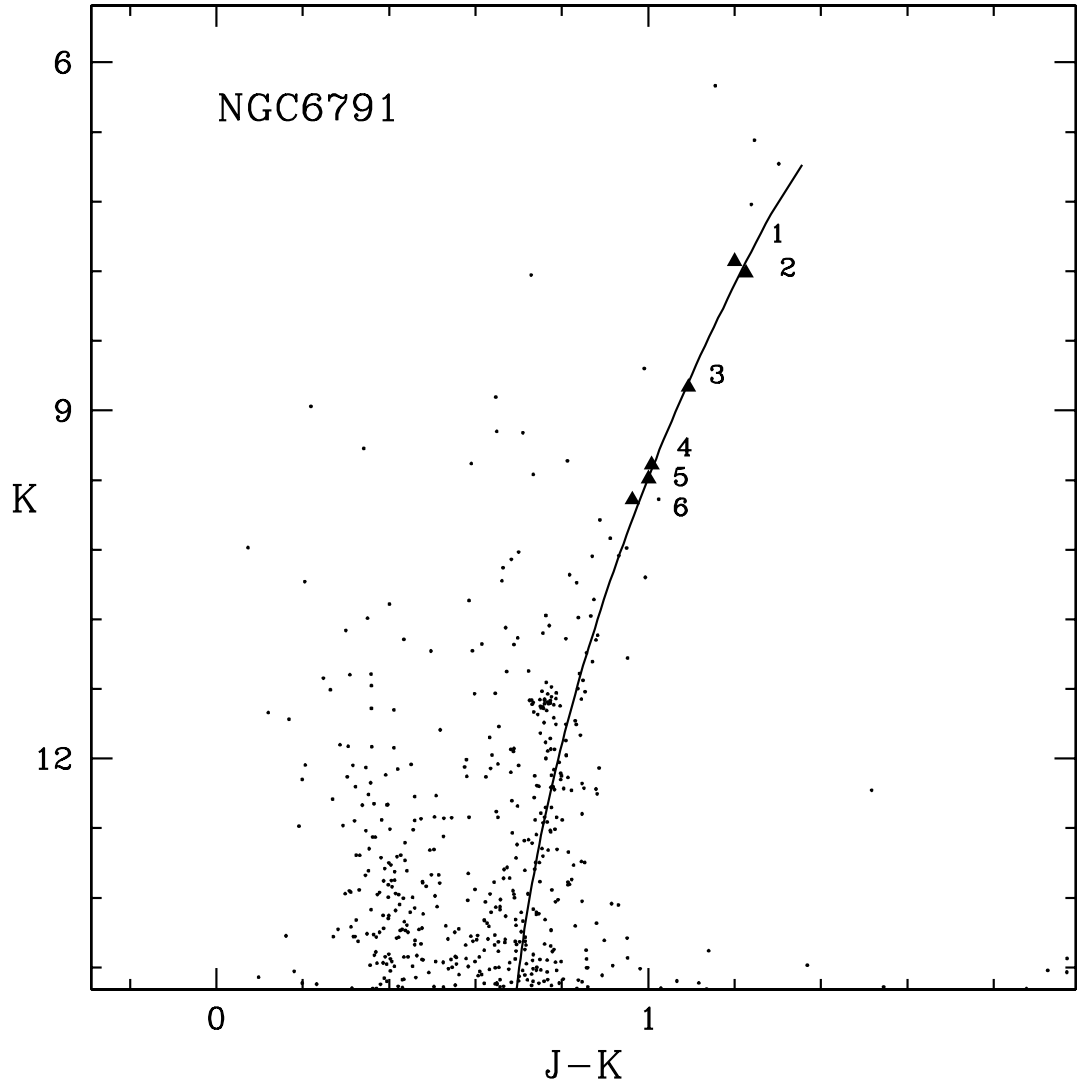}
\caption{K,(J--K) color magnitude diagram of NGC~6791 as obtained 
from 2MASS photometry.
The giant stars observed with NIRSPEC are plotted as filled triangles and the
derived RGB fiducial ridge line is superimposed as a solid line.}
\label{cmd}
\end{figure}

By using 2MASS photometry we constructed the K,(J--K) color 
magnitude diagram of NGC~6791 and 
selected 6 bright (H=9-11) giant stars (see Fig.~\ref{cmd}). 
Table~\ref{tab1} reports their 2MASS name and coordinates.
The program stars were observed at Keck on 
May 2005, with typical exposure times of 4min.   
We used NIRSPEC \citep{ml98} in the echelle mode, 
a slit width of $0\farcs43$ and a length of 12\arcsec\, 
giving an overall spectral resolution R=25,000,
and the standard NIRSPEC-5 setting, which
covers most of the 1.5-1.8 $\mu$m H-band
have been selected. 

\begin{figure*}
\epsscale{1}
\plotone{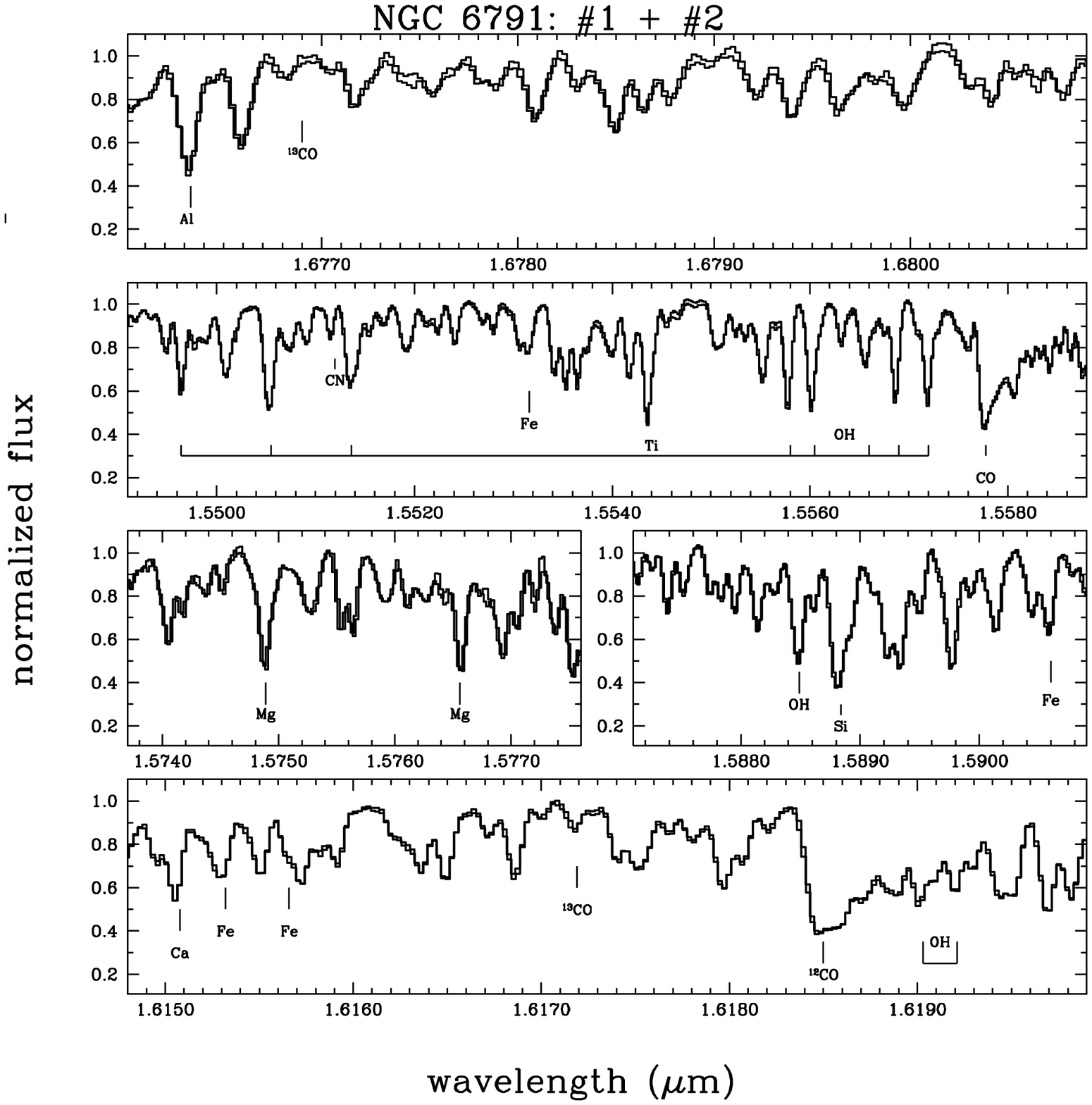}
\caption{Selected portions of the H band spectrum obtained with 
NIRSPEC for stars \#1 and \#2. 
Some features of interest are also marked.}
\label{spec}
\end{figure*}

The raw stellar spectra have been reduced using the
REDSPEC IDL-based package written
at the UCLA IR Laboratory.
Each order has been
sky subtracted by using nodding pairs and flat-field corrected.
Wavelength calibration has been performed using arc lamps and a 2-nd order
polynomial solution, while telluric features have been removed by using
a O-star featureless spectrum.
The signal to noise ratio of the final spectra is $\ge$40 and  
Fig.~\ref{spec} shows an example.

A grid of suitable synthetic spectra
of giant stars has been computed 
by varying the photospheric parameters and the
element abundances, using an updated
version of the code described in \citet{OMO93}.
By combining full spectral synthesis analysis with equivalent widths
measurements of selected lines,
we derive  abundances for Fe, C, O and other $\alpha$-elements.
The lines and analysis method have been detailed and subjected to rigorous tests in our previous studies
of  Galactic bulge field and cluster giants \citep[see][ and references therein]{ori05,ro05}.
Here we summarize the major issues.
The code uses the LTE approximation.
In the H band, most of the OH and CO molecular lines are not saturated and
can be safely treated under the LTE approximation, being roto-vibrational
transitions in the ground electronic state, providing accurate C and O 
abundances \citep{mr79,lam84,sm00}.
Detailed computations of possible NLTE effects for atomic
lines in the H band have been performed only for AlI lines 
in the Sun (see Baumueller \& Gehren (1996), finding indeed negligible 
corrections.  However,
most of the near IR atomic lines are of 
high excitation potential, indicating that they form 
deep in the atmosphere, where the LTE approximation should hold even in 
giants of low gravity.
Moreover, one of the major mechanisms which can cause a 
deviation from LTE, namely over-ionization by UV radiation, is 
less efficient in cool giants, while photon suction 
can have some relevance.
According to NLTE computations on Fe and Mg lines 
\citep[see e.g.][]{gra99,zha00} 
deviations from LTE (at a level of $\ge$0.1~dex) 
are mainly observed in stars which are significantly hotter
and more metal poor than those in our program. 
The code is based
on the molecular blanketed model atmospheres of
\citet{jbk80} in the 3000-4000~K temperature range
and the ATLAS9 models for temperatures above 4000~K.
Since in the near IR
the major source of continuum opacity is H$^-$
with its minimum near 1.6 $\mu$m,
the dependence of the results on the choice of reasonable model
atmospheres should not be critical.
However, as a check, we also computed synthetic spectra using the more updated 
NextGen model atmospheres by \citet{hau99} and we compare them with those 
obtained 
using \citet{jbk80} models, finding minor differences \citep{ro05}. 
Three main compilations of
atomic oscillator strengths are used:
the Kurucz database
(c.f. {\it http://cfa-www.harward.edu/amdata/ampdata/kurucz23/\-sekur.html}),
\citet{bg73} and \citet{mb99}.
Reference solar abundances are from \citet{gv98}.
In the first iteration, we estimate stellar temperature
from the $\rm (J-K)_0$ colors (see Table~\ref{tab2})
and the color-temperature transformation
of \citet{MFFO98} specifically calibrated on globular cluster giants.
Gravity has been estimated from theoretical evolutionary tracks,
according to the location of the stars on the RGB
\citep[see][and references therein for a more detailed discussion]{ori97}.
For microturbulence velocity an average value
$\xi$=2.0 km/s has been adopted.
More stringent constraints on the stellar parameters are obtained by the
simultaneous spectral fitting of the several CO and OH molecular bands,
which are very sensitive to temperature, gravity and microturbulence variations 
(see Figs. 6,7 of \citet{orc02}). 
The adopted values are listed in Table~\ref{tab2}.

\section{Results}

From our spectral analysis we find all the 6 stars likely members of the 
cluster, 
showing an average heliocentric radial velocity
$\rm \langle v_r \rangle = -52 \pm 1~Km/s$.
This value is in good agreement with previous estimates \citep{fj93,friel02}.
We derive abundances for Fe, C, O, Ca, Si, Mg, Ti and Al.
The final values of our best-fit models together with random 1$\sigma$ errors
are listed in Table~\ref{tab2}.
We find an average $\rm [Fe/H]=+0.35 \pm 0.02~dex$, roughly solar [$\alpha$/Fe],  
$\rm [C/Fe]=-0.35 \pm 0.03~dex$ and low $\rm ^{12}C/^{13}C\approx10\pm2$. 

\begin{figure}
\epsscale{1.0}
\plotone{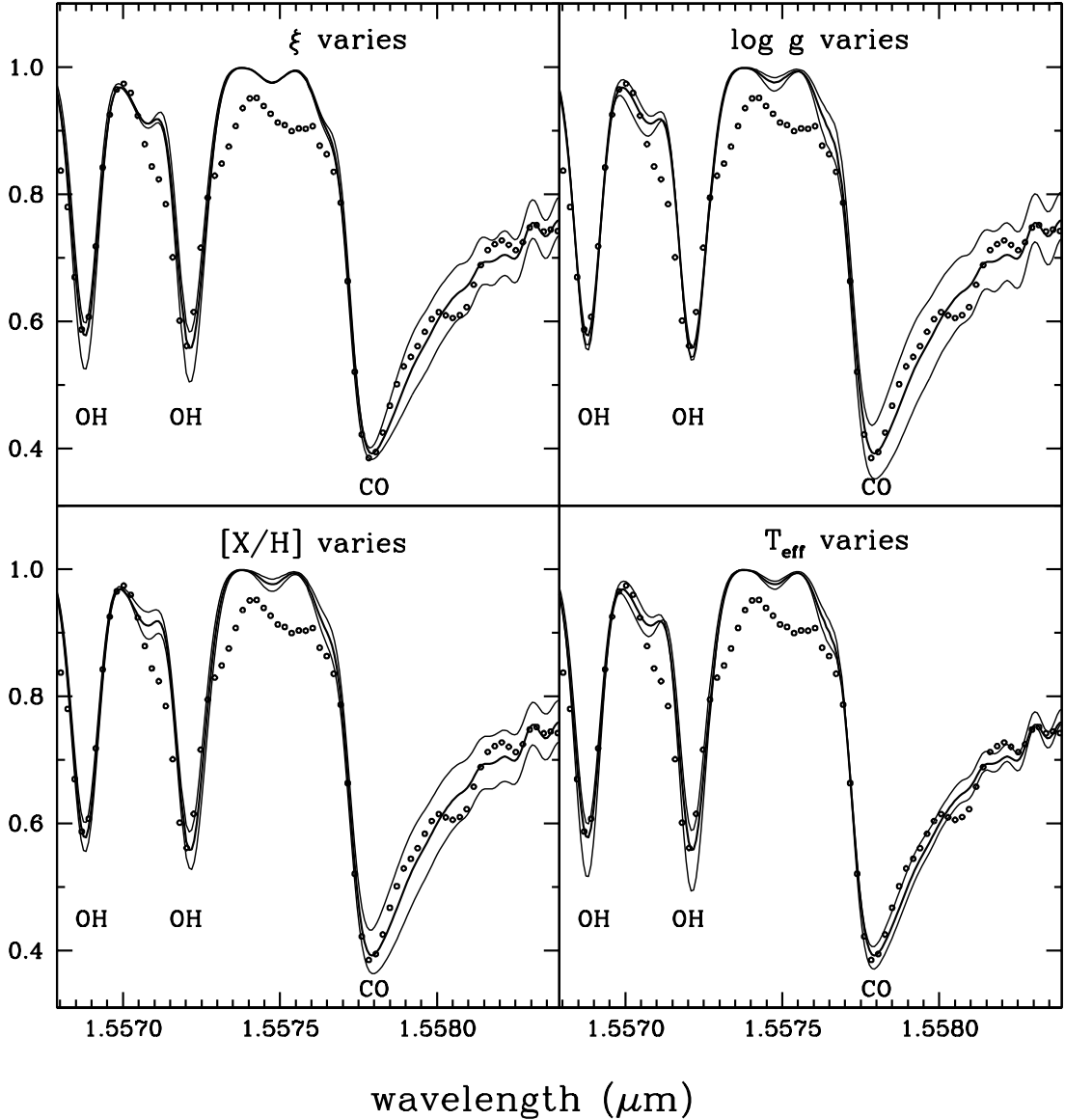}
\caption{Section of the H band spectrum of star \#3 
and our best fit (solid line), using
T$_{\rm eff}$=3600~K, log~g=1.0, $\xi$=2~km~s$^{-1}$, [Fe/H]=+0.3,
[O/Fe]=+0.0, [C/Fe]=--0.3 as reference stellar parameters 
(see also Table~\ref{tab2}).
For comparison we also plot synthetic spectra with 
different abundances and stellar parameters with respect 
to the best-fit solution. 
Bottom-left: $\rm \Delta [X/H]=\pm$0.2~dex; 
bottom-right: $\rm \Delta T_{eff}=\mp$200~K; 
top-left: $\rm \Delta \xi=\pm$0.5~km~s$^{-1}$;
top-right: $\rm \Delta log~g=\mp$0.5~dex.
}
\label{param}
\end{figure}

We also explored the results 
using models with $\rm \Delta [X/H]=\pm$0.2~dex,
$\rm \Delta T_{eff}=\pm$200~K, 
$\rm \Delta \xi=\mp$0.5~km~s$^{-1}$, 
and $\rm \Delta log~g=\pm$0.5~dex, 
with respect to the best-fit parameters.
Fig.~\ref{param} shows an example for star \#3.
It is clearly seen that models with $\pm$0.2~dex abundance 
variations give remarkably  different molecular line profiles.
Temperature variations of $\pm$200~K  
and microturbulence variation of $\pm$0.5~km/s mainly affects the OH lines, 
while gravity mainly affects the CO lines.
As a further check of the statistical significance of our best-fit solution,
we also compute synthetic spectra with
$\rm \Delta T_{eff}=\pm$200~K, $\rm \Delta log~g=\pm$0.5~dex and
$\rm \Delta \xi=\mp$0.5~km~s$^{-1}$, and with corresponding simultaneous 
variations
of the C and O abundances (on average, $\pm$0.2~dex) 
to reproduce the depth of the 
molecular features.
As a figure of merit of the statistical test we adopt
the difference between the model and the observed spectrum (hereafter $\delta$).
In order to quantify systematic discrepancies, this parameter is
more powerful than the classical $\chi ^2$ test, which is instead
equally sensitive to {\em random} and {\em systematic} errors
\citep[see also][]{ori03,ori04}.
Our best fit solutions always show $>$99\% probability
to be representative of the observed spectra, while
spectral fitting solutions with abundance variations of $\pm$0.2~dex,
due to possible systematic uncertainties of $\pm$200~K in temperature,
$\pm$0.5~dex in gravity or $\mp$0.5 km/s in
microturbulence are statistical significant at 1-2$\sigma$ level, only.
Hence, as a conservative estimate of the systematic error in the derived 
best-fit abundances,
due to the residual uncertainty in the adopted stellar parameters, one can
assume a value of $\le \pm 0.1$~dex.
However, it must be noted that since the stellar features under consideration 
show a similar trend with variations in the stellar parameters, 
although with different
sensitivities, {\it relative } abundances are less
dependent on the adopted stellar parameters (i.e. on the systematic errors)
and their values are well constrained down to $\approx \pm$0.1~dex
(see also Table~\ref{tab1}).

\section{Discussion and Conclusions}

Our derived iron abundance for NGC~6791 is in excellent agreement with the
results of \citet{pet98} and \citet{car06} and only slightly lower than the \citet{gra06} 
ones. 
All these works suggest about 0.2~dex higher iron abundances than those 
obtained by \citet{fj93,friel02}
from low\--resolution spectroscopy ([Fe/H]=+0.19~dex and [Fe/H]=+0.11~dex,
respectively).
Hence, 
the high (2-3 times solar) metallicity of NGC~6791, now confirmed by 4 independent 
surveys at medium-high resolution in the optical and in the IR, strongly supports a 
8-9 Gyr age as derived from the Main Sequence {\it Turn-Off} \citep{car06}.  

Our solar [$\alpha$/Fe] abundance ratio is in good agreement with the finding by \citet{pet98} 
for O, Ca, Ti, while their Si and Mg abundances are slightly higher.
Our [C/Fe] abundance is a factor of 2 lower than the one by \citet{pet98} but 
in good agreement with the \citet{gra06} value of [C/Fe]=-0.2.
Our and \citet{pet98} solar [O/Fe] is twice the value found by \citet{gra06}.
An overall Solar [$\alpha$/Fe] is consistent with a standard disk chemical enrichment scenario 
where both SN~II and SN~Ia contributed to the enrichment of the interstellar medium.   However,
it is interesting that the iron abundance of NGC~6791 reached +0.35 dex, more than a factor
of two greater than Solar, only a few Gyrs after the first stars were formed, relatively early in the
history of the Galaxy.  

In comparison with the Galactic bulge, NGC~6791 stars reach [Fe/H]=+0.35, only 0.15 dex lower than the
most metal rich bulge K giants reported by \citet{fmr05}.  The $\alpha$-element 
abundances are distinctly lower
than those seen in the bulge giants \citep{mr94}. In our sample of 11 bulge M giants observed with IR  
echelle spectroscopy \citep{ro05} we find [Fe/H] between 1/3 and Solar and enhanced [$\alpha$/Fe] 
abundance ratios as for K giants.
The processes that enrich the bulge rapidly and
early evidently require a star formation rate high enough to retain an alpha enhanced composition 
to nearly the Solar metallicity; this does not appear to have been the case for NGC~6791.   The age of the
Galactic bulge has been debated over the years, and ages as young as 8-9 Gyr have been discussed
seriously, especially when the luminous OH/IR stars are considered (cf. \citet{vanloon03}).   
In terms of chemistry,
there does appear to be a distinct difference between NGC~6791 and the bulge.   
The Solar [alpha/Fe] does not prove that NGC~6791 is younger than the bulge, 
but it does point to the cluster having formed well after SNe~Ia were 
able to contribute substantial iron to the interstellar medium.   
Yet another population of disk stars with similarly
high abundances are the metal rich dwarfs found in the disk \citep{castro97,pomp03}.  
These dwarfs appear to have an inner disk origin and exhibit some alpha enhancement, 
and are therefore different from NGC~6791.  Our results
would appear to indicate that the enrichment of metals is not a monotonic process in galaxies.    
A proto Milky Way 4-5 Gyr after the Big Bang had some disk regions with twice Solar metallicity.

Our low $\rm ^{12}C/^{13}C$ indicates that extra-mixing processes 
due to {\it cool bottom burning} are at work during the RGB evolution 
also at very high metallicity, confirming our findings for the metal rich giants in 
the Galactic bulge \citep[][ and references therein]{ori05}. 
  
\acknowledgments
R. Michael Rich acknowledges support from 
grants AST-0098739 and AST-0307931 from the National Science Foundation.
LO, FRF and EV acknowledge the financial support by the 
Ministero dell'Istru\-zio\-ne, Universit\`a e Ricerca (MIUR).\\
This publication makes use of data products from the Two Micron All Sky 
Survey,
which is a joint project of the University of Massachusetts and Infrared
Processing and Analysis Center/California Institute of Technology, 
founded by
the National Aeronautics and Space Administration and the National Science
Foundation.

\end{document}